\documentclass[aps,pre,onecolumn,12pt,superscriptaddress,longbibliography,nofootinbib]{revtex4-2}

\usepackage{amsmath,amsfonts,amssymb,bm,graphicx}
\usepackage{orcidlink}
\usepackage{hyperref}
\usepackage{mathrsfs}
\newcommand{\id}{\mathrm{d}}
\newcommand{\ve}{\varepsilon}
\newcommand{\cU}{\mathcal{U}}

\begin{document}

\title{Morphological instability and negative geometric friction\\ from contact with active and driven particles}

\author{Forouh Maleki \orcidlink{0009-0008-7637-6676}}
\author{Christian Maes \orcidlink{0000-0002-0188-697X}}
\email{christian.maes@kuleuven.be}
\affiliation{Department of Physics and Astronomy, KU Leuven, Belgium}

\begin{abstract}
Overdamped particles on a deformable substrate induce a geometric friction through their response to metric perturbations. Nonequilibrium dynamics couples such a perturbation to kinetic effects, which can flip that friction negative, softening the substrate's rigidity and favoring invaginations and evaginations. A breathing circle, with local curvature radius as dynamical variable, is the main example. At equilibrium the friction reduces to an ordinary potential response and stays positive. Geometric response emerges as a new route by which active and driven matter reshape their environment.
\end{abstract}

\maketitle

\section{Introduction}
The response of physical systems to external fields and forces is a cornerstone of statistical mechanics and transport theory. Much less understood is the response to perturbations of the geometry on which the particle dynamics takes place. Yet, in many physical situations, geometry is itself a dynamical degree of freedom. Examples include proteins and inclusions diffusing on biological membranes~\cite{helfrich1973elastic,Seifert1997}, active particles moving on soft interfaces and confined by walls~\cite{marchetti2013hydrodynamics,ramaswamy2010mechanics,nikola2016active}, colloids near deformable or rough substrates~\cite{bechinger2016active}, and chemical species whose transport is affected by the morphology of catalytic surfaces. In all these cases, particles move on a medium whose shape fluctuates and evolves under the influence of the particles themselves. The geometry is not merely a passive background but participates in the dynamics~\cite{prost1996shape,manneville1999activity,ramaswamy2000nonequilibrium,maitra2014activating,park2022response,king2026active,loubet2012effective,iyer2023dynamic,vutukuri2020active}.\\

Allowing the substrate geometry to evolve restores a natural form of reciprocity between matter and environment. Particles respond to the local metric, in terms of the curvature or shape of the substrate, while their motion exerts forces that in turn modify the geometry. Such reciprocal interactions are familiar in membrane physics, where inclusions generate membrane deformations~\cite{weikl1998interaction,park2022response}, and in active matter, where self-propelled particles collectively reshape interfaces and boundaries~\cite{Maleki2023,takatori2020active,le2022encapsulated,kree2026amoeboid,fily2014dynamics}.\\

The present paper focuses on the effective dynamics induced on the substrate by particle motion. Following the general idea of reduced descriptions~\cite{schilling2022coarse,widder2025generalized,zwanzig,zwanzig2001nonequilibrium,maes2014second,stef,Tanogami2022}, we regard the geometric degrees of freedom as slow variables and eliminate the particle dynamics. The resulting reduced dynamics contains an induced friction acting on the substrate, combined with white noise. The friction is called \emph{geometric}, as it is obtained from the response of the force exerted by the particles on the substrate to a weak time-dependent perturbation of the metric. This construction raises a fundamental question: how do thermal, driven, and active particles react to changes in the geometry on which they move? Beyond specific examples, the answer requires solving a largely unexplored type of linear response problem: the response of particle dynamics to metric stimuli on curved spaces. We find a nonequilibrium-induced softening and a possible morphological instability, originating from the interplay between activity and geometry.\\

The point is that, for driven and active particles, the metric perturbation has an intrinsically kinetic component, expressing the frenetic contribution~\cite{fren}, which may overwhelm the dissipative response and render the geometric friction negative. In other words, geometric response depends strongly on the nonequilibrium features of the medium: detailed balance reduces the geometric friction to the response to a potential perturbation, whereas breaking detailed balance introduces a genuinely kinetic term without equilibrium analogue, which can flip the sign. That is our main theoretical finding, with potentially important consequences for the study of morphology in the presence of active or driven matter.\\

In practice, negative geometric friction implies that the surface acquires softer modes. A positive friction stabilizes the substrate and promotes relaxation toward the reference geometry. Negative friction instead transfers energy from the nonequilibrium particle dynamics to the geometric degrees of freedom, producing a linear instability. Regions of enhanced particle density can then amplify local deformations, increase geometric fluctuations, and more easily generate invaginations or evaginations of the substrate. On biological membranes, such mechanisms may contribute to the formation of buds, protrusions, or membrane ruffles. On catalytic or chemically active surfaces, they may induce local shape changes that alter transport and reaction rates. More generally, they provide a route by which active and driven matter reshapes its environment through purely dynamical, rather than elastic, mechanisms.

\section{Setup}
\begin{figure}[t]
\centering
\includegraphics[width=0.55\linewidth]{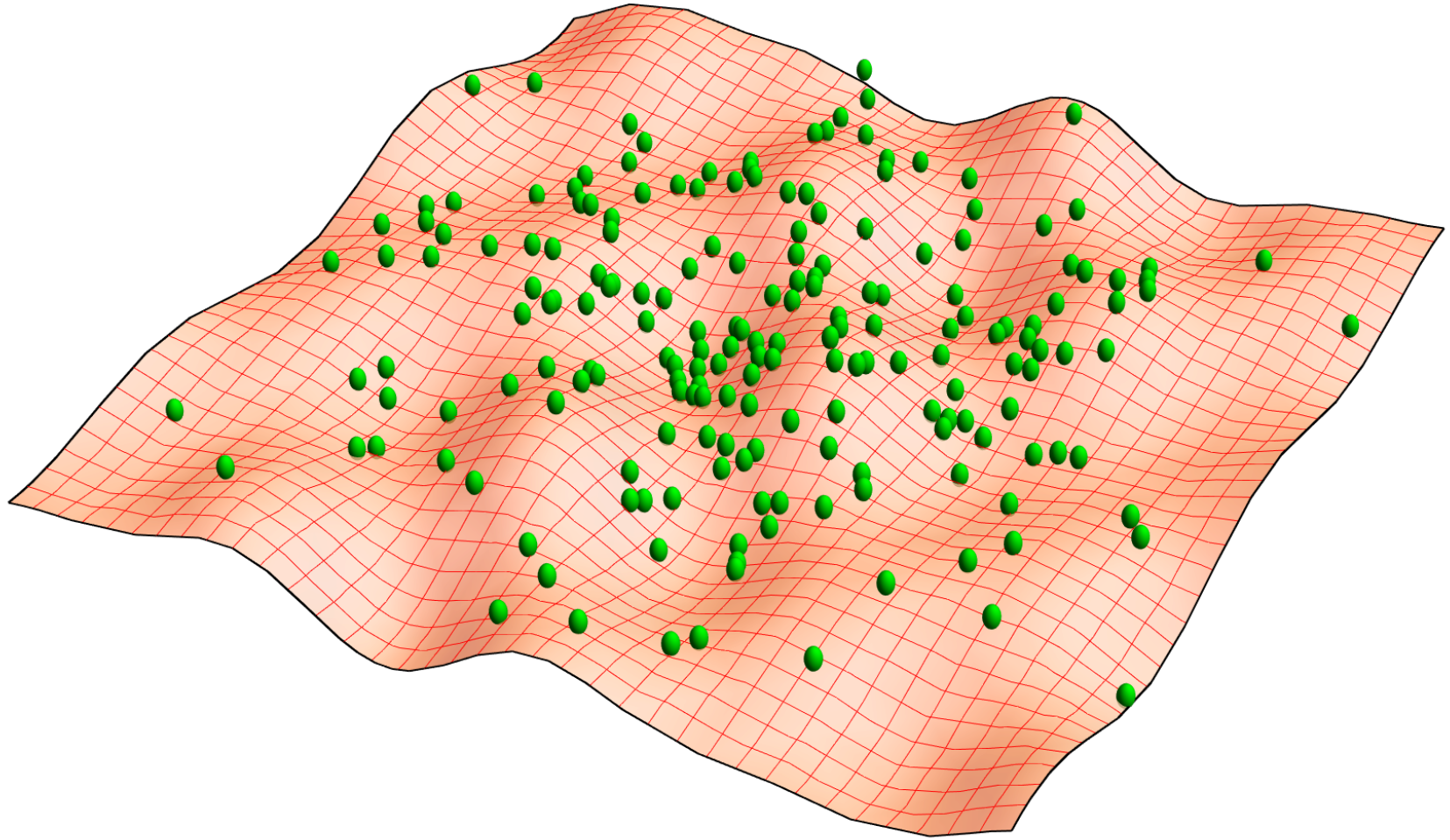}
\caption{Cartoon of a surface with particles. The metric also enters their time-symmetric dynamical fluctuations. Particles both respond to and reshape the substrate metric.}
\label{fig:surface}
\end{figure}

\begin{figure}[t]
\centering
\includegraphics[width=0.55\linewidth]{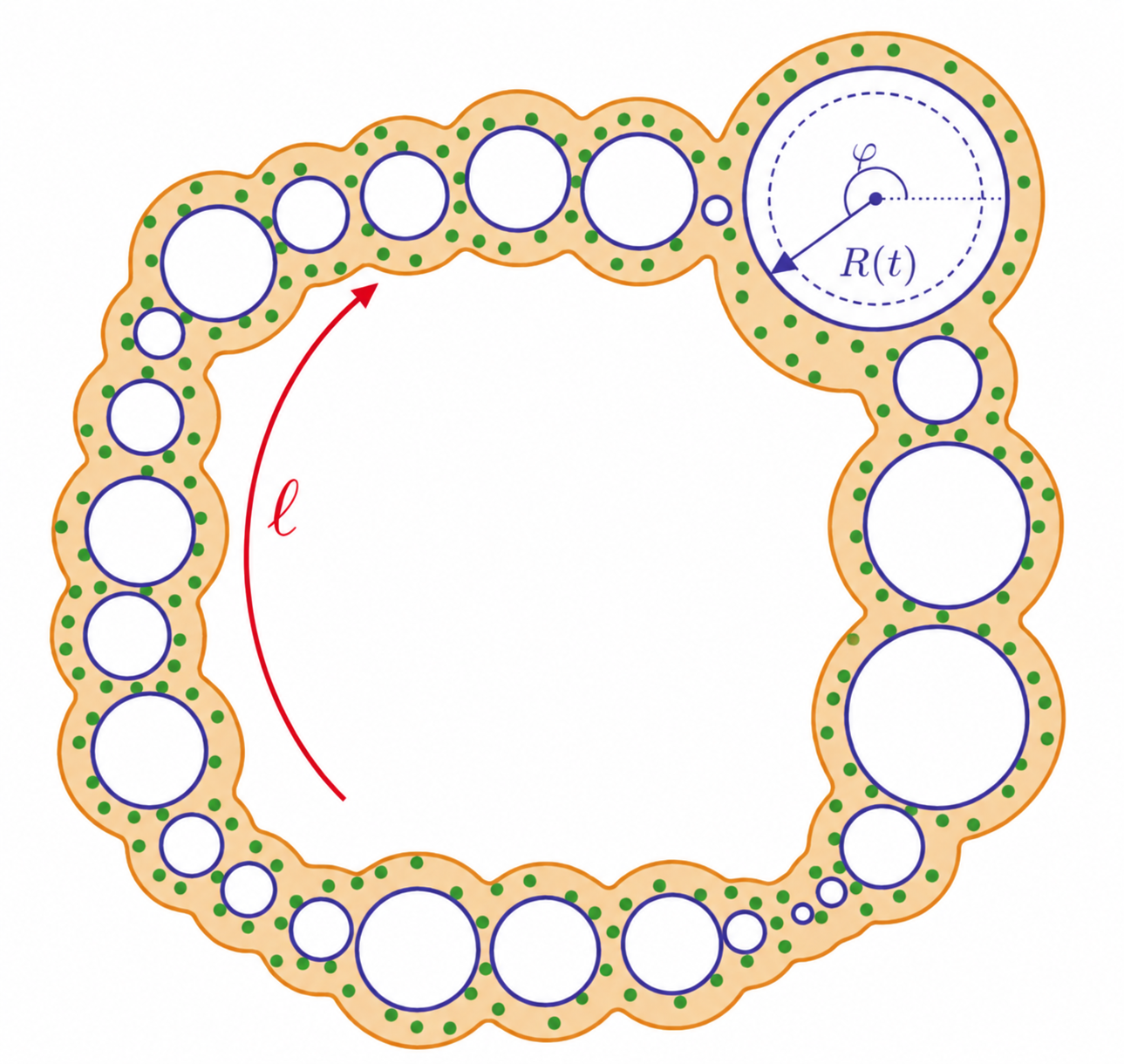}
\caption{A membrane as formed by breathing circles with radius $R_\ell(t)$. The subscript denotes the location along the membrane.  The nonequilibrium medium consists of particles surrounding and ``weighing'' on those circles.}
\label{fig:membrane}
\end{figure}

We refer to Appendix~\ref{A-sec:manifold} and Refs.~\cite{risken1989fokker,DoCarmo2016,Rosenberg1997,besse2007einstein} for the general framework of Markov diffusions on compact Riemannian manifolds; see also~\cite{nagase2026stochastic}. As illustrated in Fig.~\ref{fig:surface}, we consider a nonequilibrium medium of $N$ independent point particles with positions $X_a(t)\in(M,g(t))$, $a=1,\ldots,N$, moving on an evolving Riemannian manifold $(M,g(t))$.
In general, we start from a manifold--particle interaction or joint free energy $\lambda\,\cU(g_{ij},x)$, with coupling strength $\lambda$. The force $F$ on a geometric coordinate $q$ of the manifold (below, $q=R$) is then \cite{guven2004membrane}
\begin{equation}
F = \int_M T_{ij}\,\partial_q g^{ij}\,\id\mathrm{vol}_g,
\qquad
T_{ij} = -\frac{\lambda}{\sqrt{g}}\,\frac{\delta\cU}{\delta g^{ij}},
\label{metricdyn}
\end{equation}
and depends on the particle positions~\cite{Powers2010}. Neither the methodology nor the physical mechanism changes in general, but to isolate the geometric mechanism while keeping the analysis tractable, we specialize to an effectively one-dimensional geometry; see Fig.~\ref{fig:membrane}, also directly relevant for cylindrical and toroidal shapes, which may approximate tubes and vesicles in a biophysical context. For a general surface, the local analysis is a conjectured extension, supported by the exactly solved circular case.\\ 
The geometry is then locally represented by a slowly varying radius of curvature $R_\ell(t)$, where $\ell$ is the arclength along a curve and $t$ is time. It serves as an effective order parameter rather than a complete description of the embedding. Reduced-coordinate descriptions are commonly employed in continuum models of vesicles, budding membranes, and other soft-matter systems, where a complex surface shape is represented by a small number of collective geometric variables~\cite{JulicherSeifert1994,KhairyHoward2011,AllainBenAmar2007}. However, here we choose to ignore the interaction along the arclength, between circles at different positions $\ell$.  It will be unimportant for the calculation as long as it is not modified by the medium. For simplicity we represent that interaction with a mass term $\sim \frac{m}{2}(R-R_0)^2$ in the energy to keep each circle somewhat pinned around a reference radius $R_0$.\\

The local analysis, at fixed arclength $\ell$, is complemented by representing each particle position by its local azimuthal angle, $X_a = \varphi_a$ in Fig.~\ref{fig:membrane}.  
We are then led to an effective Helfrich-inspired free energy
\begin{equation}
\mathcal{F}[R,\{X_a\}] = \lambda\,\frac{R_0^2}{R}\sum_a G(\varphi_a) + \frac{m}{2}\,(R-R_0)^2 ,
\label{eq:Helfrich}
\end{equation}
where the instantaneous particle configuration modifies the bending energy through the ``weight'' $G$ of the particles on the circle, with dimensionless coupling $\lambda$, and where the confining term of stiffness $m$ keeps the radius near a reference curvature radius $R_0$. The geometry evolves according to the overdamped flow $\dot R = -\Gamma^{-1}\partial_R\mathcal{F}$, with bare (equilibrium) friction coefficient $\Gamma$ of the substrate. This gives the dynamics of a breathing circle,
\begin{equation}
\Gamma\,\dot R(t) = m\,\big(R_0 - R(t)\big) + \frac{\lambda R_0^2}{R^2(t)}\sum_a G(\varphi_a(t)) .
\label{eq:AllenCahn}
\end{equation}

For a circle, the embedding $Y(\varphi) = R(\cos\varphi,\sin\varphi)$ gives the induced metric $g_{\varphi\varphi} = R^2$. We write the weight in \eqref{eq:AllenCahn} as $G(\varphi) = U(\varphi)/R_0$ in terms of a potential $U$. On a circle of radius $R$, the force exerted by a particle is then
\begin{equation}
F(R,\varphi) = -\lambda\,\partial_R\cU(\varphi,R),
\qquad
\cU(\varphi,R) = U(\varphi)\,\frac{R_0}{R},
\label{infor}
\end{equation}
in agreement with \eqref{metricdyn}. Reciprocally, \eqref{eq:Helfrich} gives the conservative force in the overdamped dynamics of a particle at position $\varphi(t)$,
\begin{equation}
R(t)\,\dot\varphi (t)= -\mu\lambda\,\frac{R_0}{R^2(t)}\,\partial_\varphi U(\varphi(t)) + v_0\,\sigma(t) + \sqrt{2k_BT\mu}\;\xi(t),
\label{pdu}
\end{equation}
with mobility $\mu$, standard white noise $\xi(t)$, and propulsion term $v_0\sigma(t)$. Driven particles have $\sigma(t)\equiv+1$, and the thermal energy $k_BT=\beta^{-1}$ refers to a thermal bath in the environment. The driving force is $\mathcal{E} = v_0/\mu$, and the dissipation length is $\ell_{\rm diss} = (\beta\mathcal{E})^{-1}$; the large-driving regime has $\ell_{\rm diss}<R$. Athermal run-and-tumble particles have $T=0$ in \eqref{pdu}, and $\sigma(t)=\pm1$ flips sign at rate $\alpha$. The persistence time $\alpha^{-1}$ may, and usually does, depend on the local curvature radius $R_\ell(t)$; we write $\alpha = \alpha(R)$. The active length is $\ell_{\rm act} = v_0/\alpha$, with $\ell_{\rm act}>R$ in the ballistic regime, while the passive limit $\alpha\uparrow\infty$ keeps $\beta_{\rm eff} = 2\mu\alpha/v_0^2$ fixed as an effective inverse temperature.\\

Eqs.~\eqref{eq:AllenCahn}--\eqref{pdu} define the coupled evolution of the particles and the radius of the circle. Throughout, we consider the regime in which the geometric evolution is much slower than the particle dynamics: $\Gamma/m \gg \max\{\alpha^{-1}, R/v_0\}$ for run-and-tumble particles, and $\Gamma/m \gg \max\{R^2/(\mu k_BT), R/v_0\}$ for driven thermal particles. The particle trajectories $\{X_a(t)\}$ can then be adiabatically eliminated, yielding an effective dynamics for the curvature radius $R$. As we show below, the resulting coarse-grained manifold dynamics is in general no longer a gradient flow: nonconservative geometric forces and the associated geometric friction emerge from the nonequilibrium fluctuations of the particle medium itself, without being introduced phenomenologically. We follow the same road as for deriving activity on (initially) passive probes in driven or active media~\cite{beyen2025coupling,pei2026transferActiveMotion}.

\section{Method}
On the relaxation time scale of the medium, the manifold is quasistatic and acts as an external control parameter for the particle dynamics. For each instantaneous metric, the medium reaches the corresponding stationary distribution, while a slow evolution of the metric is regarded as a time-dependent perturbation of the medium. In that Markov approximation, the reduced manifold dynamics is characterized by a (renormalized) mean force, a friction, and a noise. For the breathing circle, after integrating out the particles from \eqref{eq:AllenCahn}, we thus expect the curvature radius to satisfy
\begin{equation}
(\Gamma + \nu)\,\dot R(t) = m\,\big(R_0 - R(t)\big) +\mathscr{S}(R(t))+ \sqrt{2B(R(t))}\, \xi(t)
\label{dynlin}
\end{equation}
see, e.g.,~\cite{pei2026transferActiveMotion,Tanogami2022,beyen2025coupling,maes2014second} for the procedure. The white noise amplitude $B$ and streaming term $\mathscr{S}$ follow in Appendix \ref{A-sec:noise},\ref{strsec}.\\

In this paper we focus on the induced geometric friction $\nu(R)$ in \eqref{dynlin}. It characterizes the linear response of the medium to the motion of the manifold. For that purpose, we compare the expected force \eqref{infor} on a slowly deforming manifold with the same force on a fixed manifold, i.e., we evaluate
\begin{equation}
\langle F(t)\rangle_\ve - \langle F(t)\rangle
\label{pero}
\end{equation}
to leading order in the local deformation $\ve$. The first expectation uses the time-dependent metric $\tilde g$; the second is the steady-state expectation for the medium dynamics \eqref{pdu} at fixed metric. The response \eqref{pero} starts from a weak perturbation of the metric,
\begin{equation}
g_{ij} \rightarrow \tilde g_{ij} = g_{ij} + \ve\,h_{ij},
\label{weakp}
\end{equation}
with $h_{ij}$ a smooth symmetric tensor field. This perturbation induces first-order corrections to all geometric quantities entering a general diffusion process, including the gradient, the Laplace--Beltrami operator, the drift field, and the forward Fokker--Planck operator. Their explicit forms are given in full generality in Appendix~\ref{A-sec:manifold}.\\

As shown in Appendix~\ref{A-sec:agarwal}, the linear response \eqref{pero} is a steady-state time correlation of the particles, given by the Agarwal-type formula~\cite{upd}
\begin{equation}
\langle F(t)\rangle_\ve - \langle F(t)\rangle
= \int_{-\infty}^t \id s\;\ve_s\left\langle \frac{L_1^\dagger\rho^{\rm stat}}{\rho^{\rm stat}}\big(\varphi(0)\big)\,;\,F\big(\varphi(t-s)\big)\right\rangle .
\label{agarwal}
\end{equation}
Here $L_1^\dagger = \partial_\ve L^\dagger_{\tilde g}\big|_{\ve=0}$ is the change of the forward generator of the particle dynamics \eqref{pdu} per unit perturbation, taken as the adjoint with respect to the unperturbed volume, and $\rho^{\rm stat}$ is the stationary particle density at fixed metric. That density is not known in general, since the particle dynamics is driven or active. We therefore work at weak coupling and compute the friction to second order in $\lambda$.\\

For the breathing circle, the perturbation is a small variation of the radius, $\ve_s =\frac{R(s)}{R(t)} - 1$, and \eqref{agarwal} reads
\begin{equation}
\langle F(t)\rangle_\ve - \langle F(t)\rangle
= -\int_{-\infty}^t \id s\;\dot R(s)\,\gamma(t-s),
\label{fric}
\end{equation}
with friction kernel $\gamma(v) = \int_v^\infty \mathcal{R}_{}(w)\,\id w$ in terms of the response function $\mathcal{R}_{}$. The geometric friction entering \eqref{dynlin} is
\begin{equation}
\nu = \int_0^\infty \gamma(v)\,\id v = \int_0^\infty u\,\mathcal{R}_{}(u)\,\id u .
\label{nudef}
\end{equation}
The detailed calculations are in Appendices~\ref{A-sec:circle} and~\ref{A-sec:friction}. A remarkable feature is that, for a thermal medium in equilibrium ($v_0=0$ in \eqref{pdu}), the metric perturbation acts through the potential only and does not change the kinetics.

\section{Results}
We find that the geometric friction is positive for an equilibrium medium but can become negative for driven and active particles. The mechanism goes via the frenetic contribution to the metric response. All expressions are to second order in $\lambda R_0/R$, with real Fourier coefficients $\tilde\cU_n$ of the potential, $\cU = \sum_n\tilde\cU_n\,e^{in\varphi}$; the friction is additive over the $N$ independent particles.\\

For driven particles ($\sigma\equiv1$ in \eqref{pdu}), the geometric friction is
\begin{equation}
\nu_{\rm dri}(R) = \frac{2N\lambda^2 R\,\beta^2}{\mu}\sum_{n>0}\frac{\mathcal{A}_n R\,(n^4-a^4) + \mathcal{B}_n\,(a^4-3n^2a^2)}{(n^2+a^2)^3},
\label{momo}
\end{equation}
where $a = R/\ell_{\rm diss} = \beta R v_0/\mu$, and
\begin{equation}
\mathcal{A}_n = \Big(\frac{\partial\tilde\cU_n}{\partial R}\Big)^2,
\qquad
\mathcal{B}_n = \tilde\cU_n\,\frac{\partial\tilde\cU_n}{\partial R} .
\end{equation}
The term $\mathcal{A}_n$ results from the dependence of the interaction potential on $R$, whereas $\mathcal{B}_n$ stems from the dependence of the derivative operator on the metric; it is a genuinely geometric (metric-induced) contribution to the dynamics. In the purely thermal case, $a=0$, only the term with $\mathcal{A}_n$ survives in \eqref{momo}: only the direct influence of the curvature on the potential counts, and thermal particles do not see the kinetic changes. For driven or active particles, the curvature enters more fundamentally.\\

\begin{figure}[t]
\centering
\includegraphics[width=0.6\linewidth]{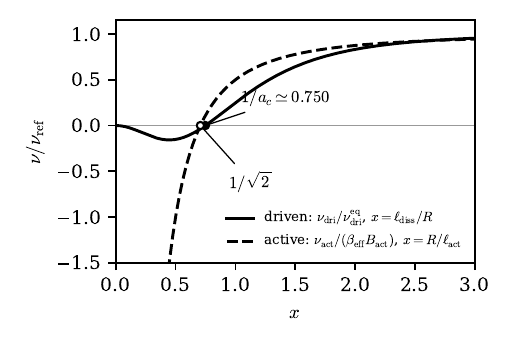}
\caption{Normalized geometric friction for $U(\varphi)=K\cos\varphi$. Solid: driven particles, $\nu_{\rm dri}/\nu_{\rm dri}^{\rm eq}$ (the value at $v_0=0$) versus $x=\ell_{\rm diss}/R$. Dashed: run-and-tumble particles with constant $\alpha$, $\nu_{\rm act}/(\beta_{\rm eff}B_{\rm act})$ versus $x=R/\ell_{\rm act}$. Small $x$ means far from equilibrium; the friction is negative for $x<1/a_c\simeq0.750$, respectively $x<1/\sqrt2$. The reduced dynamics requires $x\gg(m/\Gamma)(R/v_0)$.}
\label{fig:nu}
\end{figure}

To be explicit, we introduce the energy scale $K$ and take the first, and most relevant, mode $n=1$, $U(\varphi) = K\cos\varphi$. With $A(R) = NK^2\lambda^2R_0^2/R^2$, the geometric friction is
\begin{equation}
\nu_{\rm dri}(R) = \frac{A(R)\,\beta^2}{2\mu}\;\frac{1+3a^2-2a^4}{(1+a^2)^3},
\label{nudri}
\end{equation}
which depends on the dimensionless $a = R/\ell_{\rm diss}$, expressing the competition between the thermal energy $\beta^{-1}$ and the nonequilibrium work $Rv_0/\mu$. In equilibrium, $a=0$, and for weak driving, $\nu_{\rm dri}>0$. Beyond the threshold $a>a_c$, with $a_c = [(3+\sqrt{17})/4]^{1/2}\simeq1.334$, the geometric friction becomes negative; see Fig.~\ref{fig:nu}. For potentials of the form $\cU = U(\varphi)R_0/R$, one has $\mathcal{B}_n = -R\,\mathcal{A}_n$, and mode $n$ in \eqref{momo} changes sign only at $a = n\,a_c$: higher angular modes of the interaction require stronger driving to become negative.\\

For a medium of athermal run-and-tumble particles ($T=0$, $\alpha>0$), the friction is
\begin{equation}
\nu_{\rm act}(R) = 2N\lambda^2\mu\sum_{n>0}n^2\Big[\frac{4\alpha^2-\omega_n^2}{\omega_n^4}\,\frac{\mathcal{A}_n}{R^2} + \frac{\mathcal{B}_n}{\omega_n^2R^3} + \frac{4\alpha\alpha'}{\omega_n^4}\,\frac{\mathcal{B}_n}{R^2}\Big],
\label{nuactgen}
\end{equation}
with $\omega_n = nv_0/R$ and $\alpha' = \id\alpha/\id R$. For constant $\alpha$ and $\mathcal{B}_n = -R\,\mathcal{A}_n$, mode $n$ is proportional to $4\alpha^2 - 2\omega_n^2$; in contrast with the driven case, higher modes become negative more easily. For the same potential as above and constant $\alpha=\alpha_0$, with $\ell_{\rm act} = v_0/\alpha_0$, we obtain
\begin{equation}
\nu_{\rm act}(R) = \frac{A(R)\,\mu\,\alpha_0^2}{v_0^4}\,\Big(2 - \frac{\ell_{\rm act}^2}{R^2}\Big),
\label{RTF}
\end{equation}
where $R= R_\ell$ is the radius of a circle in Fig.~\ref{fig:membrane}. Equation~\eqref{RTF} implies negative geometric friction when $\ell_{\rm act}>\sqrt2\,R$. In both media the friction turns negative far enough from equilibrium: stronger driving shortens $\ell_{\rm diss}$, and negative friction requires $\ell_{\rm diss}<R/a_c$, whereas stronger agitation lengthens $\ell_{\rm act}$, and negative friction requires $\ell_{\rm act}>\sqrt2\,R$ (Fig.~\ref{fig:nu}).\\

Fig.~\ref{fig:nu} is physically meaningful only where the reduced Markov dynamics holds and where $\lambda R_0/R$ is small, i.e., where the medium decorrelates much faster than the radius relaxes and for weak coupling at $R\simeq R_0$. The horizontal axis $x$ can be thought as the inverse of the P\'eclet number, ratio of advective to diffusive transport. In the variables of the figure, the memory time of the medium has the same form in both cases, $\tau_{\rm med}=\tau_{\rm b}/x$, with $\tau_{\rm b}=R/v_0$ the time of one ballistic turn: $\tau_{\rm med}=R^2/(\mu k_BT)$ for driven particles, and $\tau_{\rm med}=1/\alpha$ for run-and-tumble particles, since $\alpha<v_0/R$ throughout their negative-friction regime. The condition $\tau_{\rm med}\ll\Gamma/m$ thus requires $x\gg x_{\min}=(m/\Gamma)(R/v_0)$, and negative friction is accessible in the window $x_{\min}\ll x<x_c$, with $x_c=1/a_c$ (driven) or $1/\sqrt2$ (active). That window is nonempty when $R/v_0\ll\Gamma/m$, i.e., when the particles circle the substrate many times during one relaxation of the radius. Below $x_{\min}$, where the normalized curves return to zero (driven) or diverge (active), the reduction fails; at fixed $v_0$ and $R$, the friction itself remains bounded below by the common ballistic value $-A\mu/(R^2v_0^2)$. Finally, since the relaxation time $(\Gamma+\nu)/m$ shrinks as $\nu$ becomes negative, the reduced dynamics reliably describes the softening, while the onset of instability, $\Gamma+\nu\to0$, marks the exit from the quasistatic regime.
When the tumbling rate depends on the radius as $\alpha = \alpha_0\,(R/R_0)^\eta$, Eq.~\eqref{RTF} holds with $\alpha_0$ replaced by $\alpha(R)$ and with the bracket $2(1-\eta) - \ell_{\rm act}^2/R^2$, where $\ell_{\rm act} = v_0/\alpha(R)$. Hence $\nu_{\rm act}<0$ always for $\eta\ge1$, while for $\eta<1$ the friction is negative if and only if $2(1-\eta)\,\alpha^2 < (v_0/R)^2$; see Appendix~\ref{A-sec:friction}. \\

The reduced geometric dynamics \eqref{dynlin}, the breathing of the circle, is also subject to noise. In the Markov approximation, the amplitude of the white noise is the time-integrated force autocorrelation, $B = \int_0^\infty\langle F(t)\,;\,F(0)\rangle\,\id t$. For the driven and the run-and-tumble medium, with the same potential as above,
\begin{align}
B_{\rm dri}(R) &= \frac{A}{2}\;\frac{\mu\,k_BT}{(\mu k_BT)^2 + (Rv_0)^2},\label{drib}\\
B_{\rm act}(R) &= A\,\frac{\alpha}{v_0^2}.\label{actb}
\end{align}
For a purely thermal medium ($v_0=0$ in \eqref{pdu}), the Einstein relation $\nu_{\rm dri} = \beta B_{\rm dri}$ is recovered; since $B\ge0$, the equilibrium geometric friction is positive. For the run-and-tumble medium, $\nu_{\rm act} = \beta_{\rm eff}B_{\rm act} + \nu^-_{\rm act}$, with $\beta_{\rm eff} = 2\mu\alpha/v_0^2$ and
\begin{equation}
\nu^-_{\rm act} = -\frac{A\,\mu}{R^2v_0^2},
\end{equation}
which follows the decomposition into entropic and frenetic components~\cite{beyen2025coupling}. In the passive limit $v_0,\alpha\to\infty$ at fixed $v_0^2/\alpha$, $\nu^-_{\rm act}\to0$ and the effective Einstein relation $\nu_{\rm act} = \beta_{\rm eff}B_{\rm act}$ holds.

\section{Conclusions}
Geometric friction, the response of the particle force to perturbations of the metric, separates equilibrium from nonequilibrium media in a sharp way. In equilibrium, the metric acts through the potential only, the friction is tied to the noise by the Einstein relation, and it is positive. Out of equilibrium, the metric also enters the kinetics, and that frenetic contribution can reverse the sign. For the breathing circle, driving and activity both reach the negative regime, through opposite behaviors of their characteristic lengths, within the window set by the separation of time scales. The substrate is then softened, and local invaginations and evaginations become more likely where driven or active particles are dense.\\

The calculation rests on independent point particles, weak coupling, and a circular geometry. Since the circle provides the local analysis of a surface, we expect the mechanism to carry over to general surfaces. Beyond the quasistatic regime, where $\Gamma+\nu$ approaches zero, the substrate itself enters an active geometric dynamics, in expected analogy with~\cite{pei2026transferActiveMotion,beyen2026agitated}; that regime remains to be explored. Geometric feedback and nonequilibrium driving are common to biological membranes, chemically active interfaces, and soft materials, so that metric response offers a general route by which active and driven matter reshape their environment.\\

\vspace{1cm}
\noindent {\bf Acknowledgment:} We thank Aaron Beyen for useful discussions and comments.

\section*{Data availability}
No data were created or analyzed in this study.\\
\vspace{1cm}
\appendix

\paragraph*{Overview of the Appendices.}
These appendices collect the derivations behind the results of the main text. Appendix~\ref{A-sec:manifold} recalls overdamped diffusions on a Riemannian manifold and derives the first-order change of the backward and forward generators under a perturbation of the metric. Appendix~\ref{A-sec:agarwal} gives the resulting linear response formula, of Agarwal type, and the definitions of the friction kernel and the geometric friction. Appendix~\ref{A-sec:circle} specializes to driven and active particles on a circle of radius $R$, where a slow change of the radius is the metric perturbation, and gives the stationary densities to first order in the coupling $\lambda$. Appendix~\ref{A-sec:friction} computes the geometric friction to order $\lambda^2$ for driven thermal particles and for athermal run-and-tumble particles, including a radius-dependent tumbling rate. Appendix~\ref{A-sec:noise} gives the corresponding noise amplitudes.  Appendix~\ref{strsec} gives the streaming term.  All quantities refer to a single particle; for $N$ independent particles, the friction and the noise amplitude are multiplied by $N$.

\section{Diffusions on a Riemannian manifold}\label{A-sec:manifold}

\subsection{Generators and stationary density}

The formalism of diffusions on manifolds requires some notation that we avoided in the main text. For completeness and for reference, we collect the main ingredients.

Let $(M,g)$ be a smooth, compact, connected $d$-dimensional Riemannian manifold without boundary, with metric tensor $g_{ij}$ and Riemannian volume $\id\mathrm{vol}_g = \sqrt{|g|}\,\id x^1\cdots \id x^d$. We write $\nabla$ for the Levi-Civita connection of $g$, $\nabla^i = g^{ij}\nabla_j$ for the gradient, $\operatorname{div}$ for the Riemannian divergence, and
\begin{equation}
\Delta_g f = \operatorname{div}(\nabla f) = \frac{1}{\sqrt{|g|}}\,\partial_i\Big(\sqrt{|g|}\,g^{ij}\partial_j f\Big)
\end{equation}
for the Laplace--Beltrami operator. Brownian motion on $M$ is the diffusion with generator $\frac12\Delta_g$.

We consider an overdamped diffusion $X(t)\in M$ with diffusion coefficient $D>0$ and drift vector field
\begin{equation}
b = -D\beta\,\nabla \cU + J ,
\end{equation}
where $\cU$ is a smooth potential, $\beta=(k_BT)^{-1}$ is the inverse temperature, and $J$ is a smooth non-gradient drift (a velocity field). Formally, $\dot X = b(X) + \sqrt{2D}\circ\xi$, with Stratonovich white noise $\xi$; the Stratonovich form transforms correctly under changes of coordinates. On a manifold, the process is most conveniently defined through its backward generator \cite{risken1989fokker},
\begin{equation}\label{A-L}
L f = D\,\Delta_g f + b^i\,\partial_i f = D\,\Delta_g f - D\beta\,\nabla^i\cU\,\nabla_i f + J^i\,\nabla_i f .
\end{equation}
The forward generator $L^\dagger$ is the adjoint with respect to the Riemannian volume, $\int_M \rho\, L f \,\id\mathrm{vol}_g = \int_M (L^\dagger\rho)\, f\,\id\mathrm{vol}_g$, which gives
\begin{equation}\label{A-fg}
L^\dagger \rho = -\operatorname{div}(b\,\rho) + D\,\Delta_g\rho = -\operatorname{div} j ,
\qquad
j = b\,\rho - D\,\nabla\rho = \big(-D\beta\,\nabla\cU + J\big)\rho - D\,\nabla\rho ,
\end{equation}
with $j$ the current. Since $M$ is compact and the dynamics is smooth and nondegenerate, there is a unique stationary density $\rho^{\rm stat}>0$, solving $L^\dagger\rho^{\rm stat}=0$. When $J=0$, detailed balance holds: $\rho^{\rm stat} = \rho_{\rm eq} = Z^{-1}e^{-\beta\cU}$, and the stationary current vanishes. For $J\neq 0$, the stationary current $j^{\rm stat} = b\,\rho^{\rm stat} - D\nabla\rho^{\rm stat}$ is in general nonzero.

\subsection{Metric perturbation}\label{A-sec:metricpert}

We derive the first-order change of the generators under a small deformation of the metric \cite{DoCarmo2016,Rosenberg1997,besse2007einstein}, as in Eq.~\eqref{weakp}. We write
\begin{equation}
\tilde g_{ij} = g_{ij} + \ve\, h_{ij},
\end{equation}
with $h_{ij}$ a smooth symmetric tensor field and $\ve$ a small amplitude. Throughout, indices of $h$ are raised and lowered with the unperturbed metric $g$, $h^{ij} = g^{ik}g^{jl}h_{kl}$ and $h = g^{ij}h_{ij}$, and covariant derivatives refer to $g$.

To first order in $\ve$, the geometric quantities entering the diffusion change as follows:
\begin{align}
\tilde g^{ij} &= g^{ij} - \ve\, h^{ij} + O(\ve^2), &
\id\mathrm{vol}_{\tilde g} &= \Big(1+\frac{\ve}{2}\, h\Big)\id\mathrm{vol}_g + O(\ve^2),\notag\\
\tilde\nabla^i f &= \nabla^i f - \ve\, h^{ij}\nabla_j f + O(\ve^2), &
\Delta_{\tilde g} f &= \Delta_g f + \ve\, \delta_h\Delta f + O(\ve^2),
\end{align}
with
\begin{equation}\label{A-deltaLap}
\delta_h\Delta f = -h^{ij}\nabla_i\nabla_j f - \Big(\nabla_i h^{ij} - \frac12\nabla^j h\Big)\nabla_j f .
\end{equation}
The last formula follows directly from $\Delta_{\tilde g} f = |\tilde g|^{-1/2}\partial_i(|\tilde g|^{1/2}\tilde g^{ij}\partial_j f)$, using $|\tilde g|^{1/2} = (1+\frac{\ve}{2}h)|g|^{1/2}$.

The potential may itself depend on the metric. We write $\cU_{\tilde g} = \cU + \ve\,\cU_1 + O(\ve^2)$, with first-order variation
\begin{equation}
\cU_1 = \frac{\partial\cU}{\partial g_{ij}}\, h_{ij} .
\end{equation}
Similarly, we allow a first-order change $J_1$ of the non-gradient drift. The drift then changes as $\tilde b^i = b^i + \ve\, b_1^i + O(\ve^2)$, with
\begin{equation}
b_1^i = -D\beta\,\big(\nabla^i\cU_1 - h^{ij}\nabla_j\cU\big) + J_1^i .
\end{equation}
The term $\nabla^i\cU_1$ comes from the change of the potential. The term $h^{ij}\nabla_j\cU$ comes from the change of the inverse metric used to raise the index of the gradient. The term $J_1$ is the change of the non-gradient drift. The decomposition separates the effect of the geometric deformation from that of a change in the driving.

\paragraph*{Backward generator.}
Since backward generators act on functions, they do not depend on a choice of reference measure. Writing $L_{\tilde g} = L + \ve\, L_1 + O(\ve^2)$, we obtain
\begin{eqnarray}\label{A-L1}
L_1 f &=& D\,\delta_h\Delta f + b_1^i\,\partial_i f
= D\Big[-h^{ij}\nabla_i\nabla_j f - \Big(\nabla_i h^{ij} - \frac12\nabla^j h\Big)\nabla_j f\Big]\notag\\
&&- D\beta\,\big(\nabla^i\cU_1 - h^{ij}\nabla_j\cU\big)\nabla_i f + J_1^i\,\nabla_i f .
\end{eqnarray}
The first part is the change of the diffusion through the metric. The second part is the change of the drift through the potential, the metric, and the non-conservative driving.

\paragraph*{Forward generator.}
For the linear response formula of Appendix~\ref{A-sec:agarwal}, densities must be evolved with respect to one fixed reference measure. We take the unperturbed volume $\id\mathrm{vol}_g$, and define $L_1^\dagger$ as the adjoint of $L_1$ with respect to it, $\int_M \rho\, L_1 f\,\id\mathrm{vol}_g = \int_M (L_1^\dagger\rho)\, f\,\id\mathrm{vol}_g$. Integrating by parts in \eqref{A-L1},
\begin{equation}\label{A-L1dag}
L_1^\dagger\rho = -\operatorname{div}(b_1\,\rho) + D\Big[-\nabla_i\nabla_j\big(h^{ij}\rho\big) + \nabla_j\Big(\Big(\nabla_i h^{ij} - \frac12\nabla^j h\Big)\rho\Big)\Big] .
\end{equation}
It is in divergence form, so $\int_M L_1^\dagger\rho\;\id\mathrm{vol}_g = 0$ for every $\rho$; this property is used below.

If one instead compares $L^\dagger_{\tilde g}$ and $L^\dagger_g$, each taken as the adjoint with respect to its own volume, the difference contains an extra term of order $\ve$ coming from the change of the reference measure, $\id\mathrm{vol}_{\tilde g} = (1+\frac{\ve}{2}h)\,\id\mathrm{vol}_g$. That term vanishes when $h$ is uniform, which is the case for the breathing circle of Appendix~\ref{A-sec:circle}.

\section{Linear response: the Agarwal formula}\label{A-sec:agarwal}

We take the particle dynamics in its stationary state $\rho^{\rm stat}$ for the unperturbed metric $g$, and switch on a time-dependent metric perturbation of amplitude $\ve_s$ at times $s\le t$. The generator at time $s$ is then $L + \ve_s\,L_1 + O(\ve^2)$. Starting at time $-\tau$ and to first order in $\ve$, the Duhamel (Dyson) formula gives
\begin{equation}
\langle F(t)\rangle_\ve - \langle F(t)\rangle
= \int_{-\tau}^t \id s\;\ve_s \int_M \id\mathrm{vol}_g\;\rho^{\rm stat}\,L_1\, e^{(t-s)L}F + O(\ve^2),
\end{equation}
where we used the stationarity, $\rho^{\rm stat}e^{(s+\tau)L} = \rho^{\rm stat}$. Moving $L_1$ to the density and letting $\tau\uparrow\infty$, we obtain
\begin{equation}\label{A-Ag}
\langle F(t)\rangle_\ve - \langle F(t)\rangle
= \int_{-\infty}^t \id s\;\ve_s\;\mathcal{R}_{}(t-s),
\qquad
\mathcal{R}_{}(u) = \big\langle \mathcal{Q}(X(0))\,;\,F(X(u))\big\rangle,
\qquad
\mathcal{Q} = \frac{L_1^\dagger\rho^{\rm stat}}{\rho^{\rm stat}} .
\end{equation}
The brackets denote the stationary covariance. Since $\langle \mathcal{Q}\rangle = \int_M L_1^\dagger\rho^{\rm stat}\,\id\mathrm{vol}_g = 0$, the covariance equals the correlation. Equation~\eqref{A-Ag} is the Agarwal-type formula used in the main text; we refer to \cite{upd} for its general context.

\paragraph*{Friction kernel and geometric friction.}
For the breathing circle, $\ve_s = R(s) - R(t)$, so that $\ve_t = 0$ and $\dot\ve_s = \dot R_s$. Integrating \eqref{A-Ag} by parts gives Eq.~\eqref{fric},
\begin{equation}\label{A-kernel}
\langle F(t)\rangle_\ve - \langle F(t)\rangle = -\int_{-\infty}^t \id s\;\dot R_s\;\gamma(t-s),
\qquad
\gamma(v) = \int_v^\infty \mathcal{R}_{}(w)\,\id w .
\end{equation}
For slow radius dynamics, $R(s)\simeq R(t)$ over the decay time of $\gamma$, and the induced force is $-\nu\,\dot R(t)$, with geometric friction
\begin{equation}\label{A-nudef}
\nu = \int_0^\infty \gamma(v)\,\id v = \int_0^\infty u\,\mathcal{R}_{}(u)\,\id u .
\end{equation}
The second form is how the time integrals are evaluated below.

\section{Driven and active particles on the circle}\label{A-sec:circle}

\subsection{Model and generators}

We consider a nonequilibrium medium of $N$ identical, mutually non-interacting particles on a circle of radius $R$. All derivations are for a single particle, and we drop the particle index. With $\cU(R,\varphi) = U(\varphi)\,R_0/R$, as in \eqref{infor}, the overdamped Langevin equation \eqref{pdu} for the angle $\varphi(t)$ at fixed radius $R$ reads
\begin{equation}\label{A-E1}
R\,\dot\varphi(t) = -\frac{\mu\lambda}{R}\,\partial_\varphi\cU(R,\varphi(t)) + v_0\,\sigma(t) + \sqrt{2\mu k_BT}\;\xi(t),
\end{equation}
where $\sigma(t)=\pm1$ flips sign at rate $\alpha\ge0$, $\xi(t)$ is standard white noise, and $T\ge0$ is the temperature of the environment.

The backward generator acts on functions $f(\varphi,\sigma)$ as
\begin{equation}\label{A-bwc2}
L f = \Big(-\frac{\mu\lambda}{R^2}\,\partial_\varphi\cU + \frac{\sigma v_0}{R}\Big)\partial_\varphi f + \frac{\mu k_BT}{R^2}\,\partial_\varphi^2 f + \alpha\,(\mathcal{T}-\mathcal{I})f ,
\end{equation}
where $(\mathcal{T}f)(\varphi,\sigma) = f(\varphi,-\sigma)$ is the direction flip and $\mathcal{I}$ is the identity. The forward generator is defined by $\sum_\sigma\int (L^\dagger\rho)\,f\;R\,\id\varphi = \sum_\sigma\int\rho\,(Lf)\;R\,\id\varphi$, with $\id\mathrm{vol}_g = R\,\id\varphi$:
\begin{equation}\label{A-ldag}
L^\dagger\rho = -\partial_\varphi\Big[\Big(-\frac{\mu\lambda}{R^2}\,\partial_\varphi\cU + \frac{\sigma v_0}{R}\Big)\rho\Big] + \frac{\mu k_BT}{R^2}\,\partial_\varphi^2\rho + \alpha\,(\mathcal{T}-\mathcal{I})\rho .
\end{equation}
We consider two cases:
\begin{itemize}
\item \emph{Driven thermal particles}: $\alpha=0$ and $\sigma\equiv1$, with $T>0$. The density and the generators do not depend on $\sigma$.
\item \emph{Athermal run-and-tumble particles}: $T=0$ and $\alpha>0$. Thermal noise is absent, while the $\sigma$-dependence and the tumbling term are kept.
\end{itemize}
We use the notation
\begin{equation}
a = \frac{\beta R v_0}{\mu} = \frac{R}{\ell_{\rm diss}},
\qquad
\omega_n = \frac{n v_0}{R},
\qquad
\chi_n = \frac{n^2\mu k_BT}{R^2},
\qquad
\Omega_n = \sqrt{\alpha^2-\omega_n^2} .
\end{equation}

\subsection{Perturbation by a slowly varying radius}

For a slowly evolving radius, the time dependence of $R(s)$ acts as a perturbation of the particle dynamics. The reference dynamics is \eqref{A-E1} at the fixed radius $R_t$; the actual dynamics has $R_s$, and we write $R_s = R_t + \ve_s$ with $\ve_s = R_s - R_t$ (which can be made dimensionless by dividing with $R_0$, which is not essential for the computation) small for the relevant $s\le t$. Since the circle has metric $g_{\varphi\varphi} = R^2$, this is a uniform metric perturbation, $h_{\varphi\varphi} = 2R$, and the reference-measure term discussed at the end of Sec.~\ref{A-sec:metricpert} is absent. In addition, we allow the tumbling rate to depend on the radius, $\alpha = \alpha(R)$, with $\alpha' = \id\alpha/\id R$; that is a change of the kinetics beyond the metric.

Expanding \eqref{A-bwc2} around $R_t$ gives $L = L_0 + \ve_s L_1 + O(\ve^2)$, where $L_0$ is \eqref{A-bwc2} at $R=R_t$ and $L_1 = \partial_R L$:
\begin{equation}\label{A-BO2}
L_1 f = \Big[-\mu\lambda\Big(\frac{1}{R^2}\,\partial_R\partial_\varphi\cU - \frac{2}{R^3}\,\partial_\varphi\cU\Big) - \frac{\sigma v_0}{R^2}\Big]\partial_\varphi f - \frac{2\mu k_BT}{R^3}\,\partial_\varphi^2 f + \alpha'\,(\mathcal{T}-\mathcal{I})f ,
\end{equation}
with all coefficients evaluated at $R=R_t$. Its adjoint with respect to $R_t\,\id\varphi$ is
\begin{equation}\label{A-FO2}
L_1^\dagger\rho = -\partial_\varphi\Big\{\Big[-\mu\lambda\Big(\frac{1}{R^2}\,\partial_R\partial_\varphi\cU - \frac{2}{R^3}\,\partial_\varphi\cU\Big) - \frac{\sigma v_0}{R^2}\Big]\rho\Big\} - \frac{2\mu k_BT}{R^3}\,\partial_\varphi^2\rho + \alpha'\,(\mathcal{T}-\mathcal{I})\rho .
\end{equation}
The generators for the two cases follow by setting $\alpha=\alpha'=0$, $\sigma=1$ (driven thermal particles), respectively $T=0$ (run-and-tumble particles). From now on, $R$ stands for the reference radius $R_t$.

\subsection{Stationary densities at fixed radius}

\paragraph*{Driven thermal particles.}
The stationary density solves $L^\dagger\rho^{\rm stat}_{\rm dri}=0$ with periodic boundary conditions and normalization $\int_0^{2\pi}\rho^{\rm stat}_{\rm dri}\,R\,\id\varphi = 1$. In terms of the tilted potential $\Phi(\varphi) = \lambda\cU(R,\varphi) - R v_0\varphi/\mu$, the solution is \cite{risken1989fokker,maes2008steady}
\begin{equation}\label{A-ssd}
\rho^{\rm stat}_{\rm dri}(\varphi) = \frac{1}{\mathcal{N}}\;e^{-\beta\Phi(\varphi)}\int_\varphi^{\varphi+2\pi} e^{\beta\Phi(\psi)}\,\id\psi ,
\end{equation}
with $\mathcal{N}$ fixed by the normalization. To first order in $\lambda$, $\rho^{\rm stat}_{\rm dri} = \rho^{(0)}\big(1+\lambda\, r(\varphi)\big) + O(\lambda^2)$, with $\rho^{(0)} = 1/(2\pi R)$ and
\begin{equation}\label{A-ssdriv}
r(\varphi) = \beta\, e^{a\varphi}\left[\frac{\int_0^{2\pi} e^{-a\psi}\,\partial_\psi\cU(R,\psi)\,\id\psi}{1-e^{-2\pi a}} - \int_0^\varphi e^{-a\psi}\,\partial_\psi\cU(R,\psi)\,\id\psi\right].
\end{equation}
Equivalently, $r$ is the unique periodic solution of $\partial_\varphi r - a\,r = -\beta\,\partial_\varphi\cU$, and it satisfies $\int_0^{2\pi} r\,\id\varphi = 0$. In Fourier modes,
\begin{equation}\label{A-rn}
\tilde r_n = -\frac{\beta\, i n}{i n - a}\,\tilde\cU_n ,
\end{equation}
which reduces to the Boltzmann result $\tilde r_n = -\beta\,\tilde\cU_n$ at $a=0$.\\

\paragraph*{Athermal run-and-tumble particles.}
The stationary density $\rho^{\rm stat}_{\rm act}(\varphi,\sigma)$ solves
\begin{equation}
0 = -\partial_\varphi\Big[\Big(-\frac{\mu\lambda}{R^2}\,\partial_\varphi\cU + \frac{\sigma v_0}{R}\Big)\rho^{\rm stat}_{\rm act}(\varphi,\sigma)\Big] + \alpha\big(\rho^{\rm stat}_{\rm act}(\varphi,-\sigma) - \rho^{\rm stat}_{\rm act}(\varphi,\sigma)\big),
\end{equation}
with periodic boundary conditions and normalization $\sum_\sigma\int_0^{2\pi}\rho^{\rm stat}_{\rm act}\,R\,\id\varphi = 1$. To first order in $\lambda$ \cite{beyen2025coupling},
\begin{equation}\label{A-ssa}
\rho^{\rm stat}_{\rm act}(\varphi,\sigma) = \frac{1}{4\pi R}\left[1 - \lambda\,\frac{2\alpha\mu}{v_0^2}\,\big(\cU(R,\varphi) - \bar\cU\big) + \lambda\,\sigma\,\frac{\mu}{R v_0}\,\partial_\varphi\cU(R,\varphi)\right] + O(\lambda^2),
\end{equation}
where $\bar\cU = (2\pi)^{-1}\int_0^{2\pi}\cU(R,\varphi)\,\id\varphi$. The first correction is the Boltzmann factor at the effective inverse temperature $\beta_{\rm eff} = 2\mu\alpha/v_0^2$.

\section{Geometric friction}\label{A-sec:friction}

\subsection{Response function}

We need the response function \eqref{A-Ag} with $F = -\lambda\,\partial_R\cU$, the force on the circle from Eq.~\eqref{infor}. Since $\mathcal{Q}$ and $F$ are both of order $\lambda$, the friction to order $\lambda^2$ requires $\mathcal{Q}$ to first order in $\lambda$, and only the zeroth-order ($\lambda=0$) stationary and transition densities. We have
\begin{equation}\label{A-COVc}
\mathcal{R}_{}(u) = \sum_{\sigma_0,\sigma}\int_0^{2\pi} R\,\id\varphi_0\int_0^{2\pi} R\,\id\varphi\;
\mathcal{Q}(\varphi_0,\sigma_0)\,\rho^{(0)}(\varphi_0,\sigma_0)\,\rho^0(\varphi,\sigma,u\mid\varphi_0,\sigma_0)\,F(\varphi),
\end{equation}
where $\rho^0(\varphi,\sigma,u\mid\varphi_0,\sigma_0)$ is the transition density at $\lambda=0$. For driven particles, the sums over $\sigma,\sigma_0$ are absent.

\paragraph*{Driven thermal particles.}
From \eqref{A-FO2} with $\sigma=1$, $\alpha=\alpha'=0$, and \eqref{A-ssdriv},
\begin{equation}
\mathcal{Q}_{\rm dri} = \lambda\Big[\frac{\mu}{R^2}\,\partial_\varphi^2\partial_R\cU - \frac{2\mu}{R^3}\,\partial_\varphi^2\cU + \frac{v_0}{R^2}\,r' - \frac{2\mu k_BT}{R^3}\,r''\Big] + O(\lambda^2).
\end{equation}
Using $\partial_\varphi^2 r - a\,\partial_\varphi^2 r = -\beta\,\partial_\varphi^2\cU$, the two terms proportional to $\partial_\varphi^2\cU/R^3$, which come from the radius dependence of the drift and of the diffusion coefficient, cancel. Using $\partial_\varphi^2 r = a\,r - \beta\,\partial_\varphi\cU$, we obtain
\begin{equation}\label{A-Adr}
\mathcal{Q}_{\rm dri} = \lambda\left[\frac{\mu}{R^2}\,\partial_\varphi^2\partial_R\cU + \frac{v_0}{R^2 k_BT}\,\partial_\varphi\cU - \frac{v_0^2}{\mu R\, k_BT}\, r\right] + O(\lambda^2).
\end{equation}

\paragraph*{Athermal run-and-tumble particles.}
From \eqref{A-FO2} with $T=0$ and \eqref{A-ssa},
\begin{equation}\label{A-Aac}
\mathcal{Q}_{\rm act}(\varphi,\sigma) = \lambda\mu\left[\frac{1}{R^2}\,\partial_\varphi^2\partial_R\cU - \frac{1}{R^3}\,\partial_\varphi^2\cU - 2\sigma\Big(\frac{\alpha}{R^2 v_0} + \frac{\alpha'}{R v_0}\Big)\partial_\varphi\cU\right] + O(\lambda^2).
\end{equation}
The last term comes from the radius dependence of the tumbling rate.

\paragraph*{Fourier modes.}
We write $\cU(R,\varphi) = \sum_n\tilde\cU_n(R)\,e^{in\varphi}$, so that $\tilde F_n = -\lambda\,\partial_R\tilde\cU_n$, and
\begin{align}
\tilde{\mathcal{Q}}_{{\rm dri},n} &= \lambda\left[-\frac{n^2\mu}{R^2}\,\partial_R\tilde\cU_n + \frac{i n v_0}{R^2 k_BT}\,\tilde\cU_n + \frac{v_0^2}{\mu R\,(k_BT)^2}\,\frac{i n}{i n - a}\,\tilde\cU_n\right],\label{A-Adrn}\\
\tilde{\mathcal{Q}}_{{\rm act},n}(\sigma) &= \lambda\mu\left[-\frac{n^2}{R^2}\,\partial_R\tilde\cU_n + \frac{n^2}{R^3}\,\tilde\cU_n - 2 i n\sigma\Big(\frac{\alpha}{R^2 v_0} + \frac{\alpha'}{R v_0}\Big)\tilde\cU_n\right].\label{A-Aacn}
\end{align}
We assume real Fourier coefficients, $\tilde\cU_{-n} = \tilde\cU_n\in\mathbb{R}$, as for even potentials $U(\varphi)$, and write
\begin{equation}
\mathcal{A}_n = \big(\partial_R\tilde\cU_n\big)^2,
\qquad
\mathcal{B}_n = \tilde\cU_n\,\partial_R\tilde\cU_n .
\end{equation}
The $n=0$ mode does not contribute.

\subsection{Transition densities at zero coupling}

\paragraph*{Driven thermal particles.}
At $\lambda=0$, the Fokker--Planck equation is $\partial_t\rho = -(v_0/R)\,\partial_\varphi\rho + (\mu k_BT/R^2)\,\partial_\varphi^2\rho$. With initial condition $\rho(\varphi,0) = R^{-1}\delta_{2\pi}(\varphi-\varphi_0)$, where $\delta_{2\pi}$ is the $2\pi$-periodic delta function, the Fourier modes evolve as $e^{-(i\omega_n + \chi_n)t}$, and
\begin{equation}\label{A-trdri}
\rho^0_{\rm dri}(\varphi,u\mid\varphi_0) = \frac{1}{2\pi R}\sum_n e^{-(i\omega_n+\chi_n)u}\,e^{in(\varphi-\varphi_0)} .
\end{equation}

\paragraph*{Athermal run-and-tumble particles.}
At $\lambda=0$,
\begin{equation}\label{A-acttime}
\partial_t\rho(\varphi,\sigma,t) = -\frac{\sigma v_0}{R}\,\partial_\varphi\rho(\varphi,\sigma,t) + \alpha\big(\rho(\varphi,-\sigma,t) - \rho(\varphi,\sigma,t)\big),
\end{equation}
with initial condition $\rho(\varphi,\sigma,0) = \delta_{\sigma,\sigma_0}\,R^{-1}\delta_{2\pi}(\varphi-\varphi_0)$ and periodic boundary conditions. We introduce the total density and the polarization,
\begin{equation}
p = \rho(\varphi,+,t) + \rho(\varphi,-,t),
\qquad
q = \rho(\varphi,+,t) - \rho(\varphi,-,t),
\end{equation}
with $p(\varphi,0) = R^{-1}\delta_{2\pi}(\varphi-\varphi_0)$ and $q(\varphi,0) = \sigma_0\,R^{-1}\delta_{2\pi}(\varphi-\varphi_0)$. They obey
\begin{equation}\label{A-pq}
\partial_t p = -\frac{v_0}{R}\,\partial_\varphi q,
\qquad
\partial_t q = -\frac{v_0}{R}\,\partial_\varphi p - 2\alpha\, q ,
\end{equation}
so that each of $p,q$ satisfies the telegraph equation $\partial_t^2 f + 2\alpha\,\partial_t f = (v_0/R)^2\,\partial_\varphi^2 f$. Solving mode by mode with the initial conditions above,
\begin{align}
p^0(\varphi,u\mid\varphi_0,\sigma_0) &= \frac{e^{-\alpha u}}{2\pi R}\sum_n\Big[\cosh(\Omega_n u) + \frac{\alpha - i\sigma_0\omega_n}{\Omega_n}\,\sinh(\Omega_n u)\Big]e^{in(\varphi-\varphi_0)},\notag\\
q^0(\varphi,u\mid\varphi_0,\sigma_0) &= \frac{e^{-\alpha u}}{2\pi R}\sum_n\Big[\sigma_0\cosh(\Omega_n u) - \frac{\alpha\sigma_0 + i\omega_n}{\Omega_n}\,\sinh(\Omega_n u)\Big]e^{in(\varphi-\varphi_0)},\label{A-pqTime}
\end{align}
and $\rho^0_{\rm act}(\varphi,\sigma,u\mid\varphi_0,\sigma_0) = (p^0+\sigma q^0)/2$. For $\omega_n>\alpha$, $\Omega_n = i\,|\Omega_n|$ is imaginary, and $\cosh(\Omega_n u)\to\cos(|\Omega_n|u)$, $\sinh(\Omega_n u)/\Omega_n\to\sin(|\Omega_n|u)/|\Omega_n|$; all expressions remain real and valid.

Since $F$ does not depend on $\sigma$, only $\sum_\sigma\rho^0_{\rm act} = p^0$ enters \eqref{A-COVc}. Moreover, $\mathcal{Q}_{\rm act}$ is the sum of a $\sigma_0$-independent part and a part linear in $\sigma_0$. Summing over $\sigma_0$ therefore produces only
\begin{equation}
\sum_{\sigma_0} p^0 = \frac{e^{-\alpha u}}{\pi R}\sum_n P_n(u)\,e^{in(\varphi-\varphi_0)},
\qquad
\sum_{\sigma_0}\sigma_0\, p^0 = -\frac{e^{-\alpha u}}{\pi R}\sum_n i\,S_n(u)\,e^{in(\varphi-\varphi_0)},
\end{equation}
with
\begin{equation}
P_n(u) = \cosh(\Omega_n u) + \frac{\alpha}{\Omega_n}\,\sinh(\Omega_n u),
\qquad
S_n(u) = \frac{\omega_n}{\Omega_n}\,\sinh(\Omega_n u).
\end{equation}

\subsection{Driven thermal particles}

Inserting \eqref{A-trdri} and $\rho^{(0)} = 1/(2\pi R)$ into \eqref{A-COVc}, the angular integrals give $\int_0^{2\pi}e^{i(m-n)\varphi_0}\id\varphi_0 = 2\pi\delta_{mn}$ and $\int_0^{2\pi}e^{i(\ell+n)\varphi}\id\varphi = 2\pi\delta_{\ell,-n}$, and all prefactors cancel:
\begin{equation}\label{A-Rdri}
\mathcal{R}_{\rm dri}(u) = \sum_n \tilde {\mathcal{Q}}_{{\rm dri},n}\,\tilde F_{-n}\;e^{-(i\omega_n+\chi_n)u}.
\end{equation}
The modes $\pm n$ are complex conjugates, so that $\mathcal{R}_{\rm dri}$ is real. Combining them,
\begin{equation}
\mathcal{R}_{\rm dri}(u) = \frac{2\lambda^2\mu}{R^2}\sum_{n>0} e^{-\chi_n u}\left[\Big(n^2\mathcal{A}_n - \frac{n^2a^2}{R(n^2+a^2)}\,\mathcal{B}_n\Big)\cos(\omega_n u) - \frac{n^3 a}{R(n^2+a^2)}\,\mathcal{B}_n\,\sin(\omega_n u)\right].
\end{equation}
With \eqref{A-nudef}, $\nu_{\rm dri} = \sum_n \tilde{\mathcal{Q}}_{{\rm dri},n}\tilde F_{-n}/(\chi_n+i\omega_n)^2$, which gives Eq.~\eqref{momo},
\begin{equation}\label{A-nudri}
\nu_{\rm dri} = \frac{2\lambda^2 R\,\beta^2}{\mu}\sum_{n>0}\frac{\mathcal{A}_n R\,(n^4-a^4) + \mathcal{B}_n\,(a^4-3n^2a^2)}{(n^2+a^2)^3} .
\end{equation}

\paragraph*{Example.}
For $\cU = K(R_0/R)\cos\varphi$, only $n=\pm1$ contribute, with $\tilde\cU_1 = KR_0/(2R)$, $\mathcal{A}_1 = K^2R_0^2/(4R^4)$, and $\mathcal{B}_1 = -K^2R_0^2/(4R^3)$. Then
\begin{equation}\label{A-nudri1}
\nu_{\rm dri} = \frac{\lambda^2\beta^2K^2R_0^2}{2\mu R^2}\;\frac{1+3a^2-2a^4}{(1+a^2)^3},
\end{equation}
which is positive for $a<a_c$ and negative for $a>a_c$, with $a_c^2 = (3+\sqrt{17})/4$, i.e., $a_c\simeq1.334$.

More generally, for potentials of the form $\cU = U(\varphi)\,R_0/R$, one has $\mathcal{B}_n = -R\,\mathcal{A}_n$, and the $n$th term of \eqref{A-nudri} is proportional to $n^4 + 3n^2a^2 - 2a^4$. It changes sign at $a = n\,a_c$, so that higher modes require stronger driving to become negative.

\subsection{Athermal run-and-tumble particles}

Inserting \eqref{A-pqTime}, $\rho^{(0)} = 1/(4\pi R)$ and \eqref{A-Aacn} into \eqref{A-COVc}, and splitting $\tilde {\mathcal{Q}}_{{\rm act},n}(\sigma_0)$ into its $\sigma_0$-independent part and the part linear in $\sigma_0$, we obtain
\begin{equation}\label{A-Ract}
\mathcal{R}_{\rm act}(u) = 2\lambda^2\mu\, e^{-\alpha u}\sum_{n>0} n^2\left[P_n(u)\Big(\frac{\mathcal{A}_n}{R^2} - \frac{\mathcal{B}_n}{R^3}\Big) + \frac{2}{n}\Big(\frac{\alpha}{R^2 v_0} + \frac{\alpha'}{R v_0}\Big) S_n(u)\,\mathcal{B}_n\right].
\end{equation}
With \eqref{A-nudef} and the integrals
\begin{equation}
\int_0^\infty u\,e^{-\alpha u}P_n(u)\,\id u = \frac{4\alpha^2-\omega_n^2}{\omega_n^4},
\qquad
\int_0^\infty u\,e^{-\alpha u}S_n(u)\,\id u = \frac{2\alpha}{\omega_n^3},
\end{equation}
the geometric friction is
\begin{equation}\label{A-nuact}
\nu_{\rm act} = 2\lambda^2\mu\sum_{n>0} n^2\left[\frac{4\alpha^2-\omega_n^2}{\omega_n^4}\,\frac{\mathcal{A}_n}{R^2} + \frac{1}{\omega_n^2}\,\frac{\mathcal{B}_n}{R^3} + \frac{4\alpha\alpha'}{\omega_n^4}\,\frac{\mathcal{B}_n}{R^2}\right].
\end{equation}
For potentials with $\mathcal{B}_n = -R\,\mathcal{A}_n$ and constant $\alpha$, the $n$th term is proportional to $4\alpha^2 - 2\omega_n^2$, so that, in contrast with the driven case, higher modes become negative more easily.

\paragraph*{Example.}
For $\cU = K(R_0/R)\cos\varphi$,
\begin{equation}\label{A-nuact1}
\nu_{\rm act} = \frac{\lambda^2K^2R_0^2\,\mu}{2R^6\,\omega_1^4}\Big[4\alpha^2 - 2\omega_1^2 - 4R\,\alpha\alpha'\Big],
\qquad
\omega_1 = \frac{v_0}{R}.
\end{equation}
For constant $\alpha=\alpha_0$, this is Eq.~\eqref{RTF},
\begin{equation}
\nu_{\rm act} = \frac{\lambda^2K^2R_0^2}{R^2}\,\frac{\mu\,\alpha_0^2}{v_0^4}\Big(2 - \frac{\ell_{\rm act}^2}{R^2}\Big),
\qquad
\ell_{\rm act} = \frac{v_0}{\alpha_0},
\end{equation}
which is negative for $\ell_{\rm act} > \sqrt2\,R$.

\subsection{Radius-dependent tumbling rate}
For $\alpha = \alpha_0\,(R/R_0)^\eta$, we have $R\,\alpha' = \eta\,\alpha$, and \eqref{A-nuact1} becomes
\begin{equation}\label{A-nuactm}
\nu_{\rm act} = \frac{\lambda^2K^2R_0^2\,\mu}{2R^6\,\omega_1^4}\Big[4(1-\eta)\,\alpha^2 - 2\omega_1^2\Big],
\qquad
\alpha = \alpha_0\Big(\frac{R}{R_0}\Big)^\eta .
\end{equation}
for some integer $\eta$. Hence:
\begin{itemize}
\item for $\eta\ge1$, the bracket is negative and $\nu_{\rm act}<0$ for all parameters;
\item for $\eta<1$, $\nu_{\rm act}<0$ if and only if $2(1-\eta)\,\alpha^2 < \omega_1^2$, i.e.,
\begin{equation}
2(1-\eta)\,\alpha_0^2\,R^{2\eta+2}\,R_0^{-2\eta} < v_0^2 .
\end{equation}
\end{itemize}
For instance, $\eta=0$ gives $\ell_{\rm act}>\sqrt2\,R$; $\eta=-1$ gives $4\alpha_0^2R_0^2 < v_0^2$; and $\eta=-2$ gives $6\alpha_0^2R_0^4 < R^2v_0^2$. In all cases, the geometric friction becomes negative for sufficiently large active length.

\section{Noise}\label{A-sec:noise}

In the Markov approximation, the reduced radius dynamics \eqref{dynlin} has a white noise whose amplitude is the time-integrated force autocorrelation,
\begin{equation}
B = \int_0^\infty C(u)\,\id u,
\qquad
C(u) = \big\langle F(0)\,;\,F(u)\big\rangle .
\end{equation}
The correlation $C(u)$ need not be positive at all times; it may oscillate or become negative. Its time integral, however, is nonnegative, being proportional to the spectral density of the force at zero frequency. We obtain $C(u)$ from \eqref{A-COVc} by replacing $\mathcal{Q}$ with $F$. For general real Fourier coefficients,
\begin{equation}
B_{\rm dri} = 2\sum_{n>0}\lambda^2\mathcal{A}_n\,\frac{\chi_n}{\chi_n^2+\omega_n^2},
\qquad
B_{\rm act} = 2\sum_{n>0}\lambda^2\mathcal{A}_n\,\frac{2\alpha}{\omega_n^2}.
\end{equation}
For $\cU = K(R_0/R)\cos\varphi$, with $\tilde F_{\pm1} = \lambda KR_0/(2R^2)$, this gives Eqs.~\eqref{drib}--\eqref{actb},
\begin{equation}
B_{\rm dri} = \frac{\lambda^2K^2R_0^2}{2R^2}\;\frac{\mu\,k_BT}{(\mu k_BT)^2 + (Rv_0)^2},
\qquad
B_{\rm act} = \frac{\lambda^2K^2R_0^2}{R^2}\;\frac{\alpha}{v_0^2}.
\end{equation}
For a purely thermal medium ($v_0=0$), $\nu_{\rm dri} = \beta B_{\rm dri}$, which is the Einstein relation. For run-and-tumble particles, $\nu_{\rm act} = \beta_{\rm eff}B_{\rm act} + \nu^-_{\rm act}$ with $\beta_{\rm eff} = 2\mu\alpha/v_0^2$ and, for constant $\alpha$,
\begin{equation}
\nu^-_{\rm act} = -\frac{\lambda^2K^2R_0^2}{R^2}\,\frac{\mu}{R^2v_0^2},
\end{equation}
which vanishes in the passive limit $v_0,\alpha\to\infty$ at fixed $v_0^2/\alpha$, where the effective Einstein relation is recovered.

\section{Stream}\label{strsec}
The streaming term \(\mathscr{S}(R)\) in \eqref{dynlin}, which represents the mean force exerted by the particles on the substrate, is defined as follows:
$$\mathscr{S}(R)= \int_0^{2\pi} F(R) \, \rho^{\mathrm{stat}} R\, \mathrm{d} \varphi$$
Using the stationary densities of the thermal driven and active particles, given by \eqref{A-ssd} and \eqref{A-ssa} respectively, and considering the potential $\cU = K(R_0/R)\cos\varphi$, the corresponding expressions for $\mathscr{S}(R)$ , up to second order in the coupling constant $\lambda R_0/R$ , are obtained as 
\begin{equation}
  \mathscr{S}_{\mathrm{dri}}(R) =-\dfrac{\lambda^{2} K^{2} R_{0}^{2} \,\beta}{2 R^{3}(1+a^2)} ,~~~~~~~~~~ \mathscr{S}_{\mathrm{act}}(R)=-\frac{\lambda^2 K^2R_0^2 \,\alpha\mu }{R^3\,v_0^2}
\end{equation}
We emphasize that these formul{\ae} are valid for large enough $R/R_0$ only.  This streaming term renormalizes the force on the membrane and can be removed by adding an extra term to the (bare) membrane dynamics.  It is not related to the geometric friction but note that it also depends on the nonequilibrium parameters, hence changes with respect to its equilibrium expression.

\begingroup
\linespread{1}\small

\endgroup

\end{document}